\documentclass[reprint]{revtex4-1}

\usepackage{amsmath}        
 \usepackage{lineno}
\usepackage{mathptmx}       
\usepackage{helvet}         
\usepackage{courier}        
\usepackage{type1cm}        
\usepackage{makeidx}         
\usepackage{graphicx}        
\usepackage[bottom]{footmisc}

\makeindex             

\def\go{\rightarrow}

\begin{document}


\title{Analytic Treatment of Tipping Points for Social Consensus in Large Random Networks}

\author{W. Zhang, C. Lim}
\email{\{zhangw10,limc\}@rpi.edu}
\affiliation{NeST Center and Mathematical Sciences Dept., Rensselaer Polytechnic Institute, 110 8th Street, Troy, New York 12180-3590, USA}


\author{B.K. Szymanski}
\email{corresponding author: szymansk@cs.rpi.edu}
\affiliation{NeST Center \& Computer Science Dept., Rensselaer Polytechnic Institute, 110 8th Street, Troy, New York 12180-3590, USA}

\begin{abstract}
We introduce a homogeneous pair approximation to the Naming Game (NG) model
by deriving a six-dimensional ODE for the two-word Naming Game.
Our ODE reveals the change in dynamical behavior of the Naming Game as a
function of the average degree $<k>$ of an uncorrelated network. This result
is in good agreement with the numerical results. We also analyze the extended
NG model that allows for presence of committed nodes and show that there is a
shift of the tipping point for social consensus in sparse networks.
\end{abstract}

\maketitle


\section{Introduction}

The dynamics of social influence has been heavily studied in the
network science literature
\cite{Schelling1978,Castellano2009, Kempe2003}. Some of the
models used include the Voter Model \cite{vote}, the threshold model
\cite{Granovetter1978}, the Bass model \cite{Bass1969} and the
Naming Game models \cite{Steels1995,Steels1998,Baronchelli2005,Baronchelli2006}.
The last model is the focus of our paper. In this model, each node is assigned a list of names as its opinions chosen from an alphabet $S$. In each time step, two
neighboring nodes, one listener and one speaker are randomly picked.
The speaker randomly picks one name from its name list and sends it
to the listener. If the name is not in the list of the listener, the
listener will add this name to its list, otherwise the two
communicators will achieve an agreement, i.e. both collapse their
name list to this single name. The variations of this game can be
classified as the ``Original'' (NG), ``Listener Only'' (LO-NG) and
``Speaker Only'' (SO-NG) types \cite{Baronchelli2} regarding the
update when the communicators make an agreement, and as the
``Direct'', ``Reverse'' and ``Neutral'' types regarding the way
that the two communicators are randomly picked, as defined in \cite{DallAsta2006}. These variations
have different behaviors but can be analyzed in the similar way. In
this paper we mainly focus on the ``Original'' ``Direct'' version using binary alphabet of two symbols, denoted here as A and B.

A key feature in the Naming Game models (symmetric binary agreement version) is
that once a susceptible individual adopts the new opinion, he can
still revert back to his old opinion at all subsequent times which
is suited to studying the dynamics of competing opinions where
switching one's opinion has little overhead, and where the opposing
opinions A and B are not socially, culturally or morally ranked.
These dynamics correspond to a particular case of the 2-convention
NG introduced in \cite{Castellano2009} with the trust parameter
$\beta = 1$. Other versions of the Naming Games have been developed
that address the issues of overall social preference of one opinion
over the other through an asymmetry in the stickiness of each
opinion.

Another significant difference between the Naming Games
(NG) and other stochastic games on networks including population
genetics models is the symmetric forms of the NG do not take into
account the selective survivability or fitness of agents that adopt
one opinion over another.

A third significant difference between the NG and other social
influence models such as the Voter models \cite{vote} is that in the
NG, an agent is allowed to hold more than one opinions before
switching to the other opinion. This changes the expected time to
consensus starting from uniform initial conditions even in perfectly
symmetric form of the models. Numerical studies in
\cite{Castellano2009} have shown that for the symmetric NG on a
complete graph, starting from the state where each agent has one of
the two opinions with equal probability, the system first achieves
consensus in order $\ln N$ number of time steps, as compared to
order $N$ time steps for the Voter models. Here $N$ is the number
of nodes in the network, and unit time consists of $N$
speaker-listener interactions.

Numerical studies in
\cite{Castellano2009} have shown that for the symmetric NG on a
complete graph, starting from the state where each node has one of
the two opinions with equal probability, the number of time steps
needed by the system to achieve consensus is of the order $\ln N$,
while the number of time steps to consensus in the Voter models is
of the order $N$. Here $N$ is the number
of nodes in the network, and a single time step consists of $N$
speaker-listener interactions.

In this paper, we also address the nearly-symmetric cases of the Naming Game models
where a single asymmetry is introduced into the models through the
random inclusion of a minority fraction of committed agents whose
opinion are fixed for all times to be A, say. The key observable is
the expected time to consensus of the A opinion and its dependence
on (I) the committed fraction and (II) the network topology.

In \cite{Zhang11} and \cite{Xie11}, we show that, for a complete
graph, when the committed fraction grows beyond a critical value
$p_c \approx 0.0979$, there is a super-exponential decrease in the
time taken for the entire network to adopt the A opinion.
Specifically, using a straight forward mean field approach,
coarse-grained stochastic analysis, and direct simulations of the
NG, we show that for $p < p_c$, the mean consensus time $T_c \sim
e^N$, while for $p
> p_c$, $T_c \sim \ln N$ . In the presence of
committed agents of opinion $A$, the only absorbing state in the
associated random walk Markov chain model \cite{Zhang11} is the
consensus state of opinion $A$ while the near-consensus state where
all susceptible agents have the $B$ opinion becomes a reflecting
state. Similarly, the averaged or mean field system of two coupled
nonlinear differential equations \cite{Xie11}, undergoes a
saddle-node bifurcation when $p = p_c$, in which the saddle point
(symmetric in phase plane in the case with no committed agents)
merges with a node to form a new equilibrium point of saddle-node
type \cite{Strogatzbook} \cite{Golubitskybook}. The $T_c \sim \ln N$
time scale comes from the slow dynamics along the center manifold
between the saddle-node and the consensus of $A$ opinion, and is the
same order as the symmetric case where there are no committed
agents. In contrast, for $p < p_c$, the $T_c \sim e^N$ time scale is
due to the additional numerous time steps spent along that portion
of the center manifold which is between the stable fixed point and
the saddle-point where the latter corresponds to a state with a
larger fraction of agents of opinion $A$ than the former.

Simulations on a range of sparse random networks with 100 - 10000
nodes have shown, after extensive and costly numerical experiments,
that the above tipping point effect of the NG with a minority
fraction of committed agents is a very robust phenomenon with
respect to underlying network topology. Of particular significance
is the numerical / empirical observation \cite{Xie11} that as one
lowers the average degree of the underlying random network, the
tipping fraction $p_c$ decreases.

In this paper, we analytically establish the numerical discovery using a refined mean field
approach \cite{Zhangcomple12} and report on precise changes in NG
dynamics with respect to the average degree $<k>$ of an uncorrelated
underlying network which are beyond the reach of the straight
forward mean field model in \cite{Zhang11}, \cite{Xie11}.
Specifically, the critical tipping fraction in the binary agreement
model decreases to a minimum of 5 percent when the average degree
$<k> = 4$ from a maximum of 10 percent for complete graphs. This
shows that the new mean field model is in better agreement with the
numerical results reported above in \cite{Xie11} and provides a much
improved approximation to NG dynamics on large random networks in
comparison to the straight forward mean field model in \cite{Xie11}.

\section{Improved Mean Field Approach}

Although the basic mean field approach applied to the NG models
\cite{Xie11} have yielded significant results such as a phase
transition at a critical fraction of the committed agents in the
network, the tipping point \cite{Xie11}, its theoretical predictions
deviate from the results of simulations on complex networks
especially when the network is relatively ``sparse''. The
qualitative changes in dynamical behavior of the network under
social games such as the NG, in terms of its average degree or the
degree distribution, is important in network science, and we report
here the significant results of a refined mean field model for the
NG with committed agents.

Recently, a so-called homogeneous pair approximation has been
introduced to study the dynamics of the voter model \cite{Vazquez}\cite{Pugliese},
a model simpler than the NG, which improves the basic mean field
approximation by taking account of the correlation between the
nearest neighbors. Their analysis is based on the master equation of
the active links, the links between nodes with different opinions.
Although it shows a spurious transition point, it captures most
features of the dynamics and works very accurately on most
uncorrelated networks such as Erdos-Renyi (ER) and scale-free (SF)
networks.

In this paper, we apply a similarly improved mean field approach to
the NG, especially the binary agreement model. Compared to the voter
model, there are more than one type of active links (edges) in the
NG, so we have to analyze all types of links including active and
inert ones. As a consequence, instead of a one dimensional averaged
nonlinear ODE in the voter model, we have a six dimensional
nonlinear coupled system. We will derive the equations by analyzing
all possible updates in the process and write it in a matrix form
with the average degree $<k>$ as a explicit parameter. In contrast
to the basic mean field theory, this improved ODE approximation
clearly shows how the NG dynamics changes when $<k>$ decrease to 1,
the critical value for ER network to have giant component, and
converges to the basic mean field equations in \cite{Xie11} when
$<k>$ grows to infinity. Next we show the significantly better
agreement between the theoretical predictions of the new mean field
theory and the simulations on ER network. Using this improved model,
we are able to predict and replicate the empirically observed
lowering of critical tipping fraction in low average degree
networks, i.e. we need fewer committed agents to force a global
consensus in a loosely connected social network.

\section{The Model}
In the ``Original'' ``Direct'' version of Naming Game, every agent in the network has a naming list in its memory. In each time step, a speaker is randomly picked first and then a listener is randomly picked from the speaker's neighbors (this order is called ``Direct''). The speaker picks one name from its memory and sends it to the listener. If the listener does not know this name , it adds this new name to its list. Otherwise, both agents delete all names in their list but the one sent (this update is the ``Original'' version).

Consider the NG dynamics on an uncorrelated random network where the
presence of links are independent, together with the following
assumptions which comprise the foundation of the homogeneous pair approximation:\\
\begin{enumerate}
\item The opinions of direct neighbors are correlated, while
there is no extra correlation besides that through the nearest neighbor.
To make this assumption clear, suppose three nodes in the network are
linked as 1-2-3 (so there is no link between 1 and 3). Their opinions
are denoted by random variables $X_1$,$X_2$,$X_3$, correspondingly.
Therefore our assumption says: $P(X_1|X_2)\neq P(X_1)$, but
$P(X_1|X_2,X_3)=P(X_1|X_2)$. This assumption is valid for all
uncorrelated networks (Chung-Lu type network \cite{Chung}, especially
the ER network).
\item The opinion of a node and its degree are mutually independent.
Suppose the node index $i$ is a random variable which labels a
random node. In terms of the opinion and degree of node $i$,
denoted respectively as $X_i$ and $k_i$, this assumption means
$E[k_i|X_i]=<k>$, $P(X_i|k_i)=P(X_i)$ and $P(X_i|X_j,k_i,k_j)=P(X_i|X_j)$
where $j$ is a neighbor of $i$. This assumption is obviously satisfied
for the networks in which every node has the same degree (regular
geometry), but it is also valid for the network whose degree distribution
is concentrated around its average (for example, Gaussian distribution
with relatively small variance or Poisson distribution with not too
small $<k>$). It can be shown that this assumption is good enough for ER
network.
\end{enumerate}

In other words, in typical mean field language, the probability
distribution of the neighboring opinions of a specific node is an
effective field. This field is however not uniform over the network
but depends only on the opinion of the given node. For an
uncorrelated random network with N nodes and average degree $<k>$,
the number of links in this network is $M=N<k>/2$. We denote the
numbers of nodes taking opinions A,B and AB as $n_A$, $n_B$,
$n_{AB}$, their fractions as $p_A$,$p_B$, $p_{AB}$. We also denote
the numbers of different types of links as
$\vec{L}=[L_{A-A},L_{A-B},L_{A-AB},L_{B-B},L_{B-AB},L_{AB-AB}]^T$,
and their fractions are given by $\vec{l}=\vec{L}/M$. We take
$\vec{L}$ or $\vec{l}$ as the coarse grained macrostate vector. The
global mean field is given by:
\begin{eqnarray*}
\vec{p}(\vec{L})&=&\left(\begin{matrix}  p_A \\
                                p_B \\
                                p_{AB}\\
                               \end{matrix}\right)={1\over 2M}\left(
                               \begin{matrix}  <k>n_A \\
                                                <k>n_B \\
                                                <k>n_{AB}\\
                                                \end{matrix}\right) \\
&=&{1\over 2M}\left(\begin{matrix}
                                                                              2L_{A-A}+L_{A-B}+L_{A-AB} \\                                                                                 L_{A-B}+2L_{B-B}+L_{B-AB} \\                                                                                 L_{A-AB}+L_{B-AB}+2L_{AB-AB}\\
                                                                                 \end{matrix}\right).\\
\end{eqnarray*}

Suppose $X_i$, $X_j$ are the opinions of two neighboring nodes. We simply write $P(X_i=A|X_j=B)$, for example, as $P(A|B)$. We also represent the effective
fields for all these types of node in terms of $\vec{L}$:
{\small $$\overrightarrow{P(\cdot|A)}(\vec{L})=\left(\begin{matrix}  P(A|A) \\
                                P(B|A) \\
                                P(AB|A)\\
                               \end{matrix}\right)={1\over 2L_{A-A}+L_{A-B}+L_{A-AB}}\left(\begin{matrix}  2L_{A-A}\\
                                L_{A-B} \\
                                L_{A-AB}\\
                               \end{matrix}\right),$$

$$\overrightarrow{P(\cdot|B)}(\vec{L})=\left(\begin{matrix}  P(A|B) \\
                                P(B|B) \\
                                P(AB|B)\\
                               \end{matrix}\right)={1\over L_{A-B}+2L_{B-B}+L_{B-AB}}\left(\begin{matrix}  L_{A-B}\\
                                2L_{B-B} \\
                                L_{B-AB}\\
                               \end{matrix}\right),$$

$$\overrightarrow{P(\cdot|AB)}(\vec{L})=\left(\begin{matrix}  P(A|AB) \\
                                P(B|AB) \\
                                P(AB|AB)\\
                               \end{matrix}\right)={1\over L_{A-AB}+L_{B-AB}+2L_{AB-AB}}\left(\begin{matrix}  L_{A-AB}\\
                                L_{B-AB} \\
                                2L_{AB-AB}\\
                               \end{matrix}\right).$$}

To derive the averaged nonlinear ODE for NG dynamics, we calculate
the expected change of $\vec{L}$ in one time step, $E[\Delta \vec{L}|\vec{L}]$. In the following equation, we add up the expectation $E[\Delta \vec{L}|\vec{L},\omega]$ conditioned by each type of nodes communications ($\omega$), and weighted by the probability of this type of nodes communications, $P(\omega)$.

\begin{equation}
E[\Delta \vec{L}|\vec{L}]=\sum_{\omega}P(\omega)E[\Delta \vec{L}|\vec{L},\omega]. \label{eq1}
\end{equation}

For brevity, we display the calculation of one term in the above summation as example. Consider the case: listener holds opinion A while
speaker has opinion B, and denote this case by $\omega=(B\go A)$. The probability for this type of
communication is
$$P(B\go A)=p_B P(A|B)={1\over 2M}L_{A-B}.$$
 The direct consequence of this communication is that the link between the listener and speaker changes from A-B into AB-B, so $L_{A-B}$ decreases by 1 and
$L_{B-AB}$ increases by 1. This \textbf{\emph{direct change}} of $\vec{L}$ is represented by
$$\vec{D}(B\go A)=\left(\begin{array}{c} 0\\
                                                          -1\\
                                                          0\\
                                                          0\\
                                                          1\\
                                                          0\\
                               \end{array}\right).$$

Furthermore, since the listener changes
opinions from A to AB, all his other related links change. The
number of these links is on average $<k>-1$ (here we use the assumption 2, $E[k_i|X_i]=<k>$.). The probabilities for each link to
be A-A, A-B, A-AB before the communication is given by
$\overrightarrow{P(\cdot|A)}$ (here we use assumption 1). After the
communication, these links will change into AB-A, AB-B, AB-AB
correspondingly. This \textbf{\emph{related change}} of $\vec{L}$ is represented by

$$(<k>-1)\left(\begin{array}{ccc}  -1 & 0 & 0 \\
                                  0 & -1 & 0 \\
                                  1 & 0 & -1 \\
                                  0 & 0 & 0 \\
                                  0 & 1 & 0 \\
                                  0 & 0 & 1 \\
                               \end{array}\right)\left(\begin{array}{c}
                                                   P(A|A)\\
                                                   P(B|A)\\
                                                   P(AB|A)\\
                               \end{array}\right).$$

The 6-by-3 matrix in the above expression indicates the link correspondence between A-A, A-B, A-AB and AB-A, AB-B, AB-AB when a ``A node" changes into ``AB node", we denote it by

$$Q_A=\left(\begin{array}{ccc}  -1 & 0 & 0 \\
                                  0 & -1 & 0 \\
                                  1 & 0 & -1 \\
                                  0 & 0 & 0 \\
                                  0 & 1 & 0 \\
                                  0 & 0 & 1 \\
                               \end{array}\right).$$

Let $\vec{R}(B\go A)=Q_A\overrightarrow{P(\cdot|A)}$, we obtain:
$$E[\Delta\vec{L}|\vec{L},B\go A]=\vec{D}(B\go A)+(<k>-1)\vec{R}(B\go A).$$

On the right hand side of the above equation, the first term represents the \textbf{\emph{direct change}} and the second term represents the \textbf{\emph{related change}}.

Similarly, we analyze all the other terms in equation (\ref{eq1}) for different $\omega$
(the listener and speaker's opinions), and write the weighted sum in matrix form, we obtain:
$$E[\Delta\vec{L}|\vec{L}]={1\over M} \left[D+(<k>-1)R\right]\vec{L},$$

where $D$ is a constant matrix whose column vectors come from linear combinations of $\vec{D}(\omega)$'s:

$$D=\left(
      \begin{array}{cccccc}
        0 & 0 & {3\over 4} & 0 & 0 & {1\over 2} \\
        0 & -1 & 0 & 0 & 0 & 0 \\
        0 & {1\over 2} & -1 & 0 & 0 & 0 \\
        0 & 0 & 0 & 0 & {3\over 4} & {1\over 2} \\
        0 & {1\over 2} & 0 & 0 & -1 & 0 \\
        0 & 0 & {1\over 4} & 0 & {1\over 4} & -1 \\
      \end{array}
    \right),$$

and matrix $R$ is a function of $\vec{L}$, given by column vectors which come from $\vec{R}(\omega)$'s:

\begin{eqnarray*}
R=\big(\vec{0}, {1\over 2}[Q_A \overrightarrow{P(\cdot|A)}+Q_B \overrightarrow{P(\cdot|B)}],
Q_A [{1\over 4}\overrightarrow{P(\cdot|A)}-{3\over 4}\overrightarrow{P(\cdot|AB)}],\\
\vec{0},Q_B[{1\over 4}\overrightarrow{P(\cdot|B)}-{3\over 4}\overrightarrow{P(\cdot|AB)}] , -(Q_A+Q_B)\overrightarrow{P(\cdot|AB)}\big).
\end{eqnarray*}

Here $Q_B$, similar to $Q_A$ defined above, indicates the link correspondence between B-A, B-B, B-AB and AB-A, AB-B, AB-AB when a ``B node'' changes into ``AB'',
$$  \ Q_B=       \left( \begin{array}{ccc}
          0 & 0 & 0 \\
          -1 & 0 & 0 \\
          1 & 0 & 0 \\
          0 & -1 & 0 \\
          0 & 1 & -1 \\
          0 & 0 & 1 \\
        \end{array}\right).
$$

When ``AB node'' changes into ``A'' or ``B'', the link correspondence is given by $-Q_A$ or $-Q_B$ respectively.

Then we normalize $\vec{L}$ by the total number of links $M$ and
normalize time by the number of nodes $N$ to obtain:

\begin{eqnarray}
{d\over dt}\vec{l}&=&{N\over M} E[\Delta \vec{L} |\vec{L}] ={N\over M}\left[D+(<k>-1)R\right]\vec{l} \nonumber\\
                  &=&2\left[{1\over <k>}D+({<k>-1\over <k>})R\right]\vec{l}.
                  \label{eq2}
\end{eqnarray}
Thus, we derived the new mean field ODEs for $\vec{l}$ and the
average degree $<k>$ of the underlying social network on which the
NG is played is explicit in the formula. In the last line, the first
term is linear and comes from the \textbf{\emph{direct change}} of the link between the
listener and the speaker. The second term is nonlinear and comes
from the \textbf{\emph{related changes}}.

Under the previous basic mean field assumptions in \cite{Xie11}, the
first term does not exist because there is no specific ``speaker''
and every one receives messages from the effective mean field. When
$<k>\go 1$, the new ODE becomes:
$${d\over dt}\vec{l}=2D\vec{l},$$
which is a linear system. When $<k>\go \infty$, this ODE becomes:
$${d\over dt}\vec{l}=2R\vec{l}.$$
If in matrix $R$ we further require
$\overrightarrow{P(\cdot|A)}=\overrightarrow{P(\cdot|B)}=\overrightarrow{P(\cdot|AB)}=\vec{p}$
and transform the coordinates by $\vec{L}\go \vec{p}(\vec{L})$, this
ODE reverts to the one we have under the basic mean field assumption
in \cite{Xie11}.

\section{Numerical Results without committed agents}
\begin{figure}[!hbp]
  \includegraphics[width=0.55\textwidth]{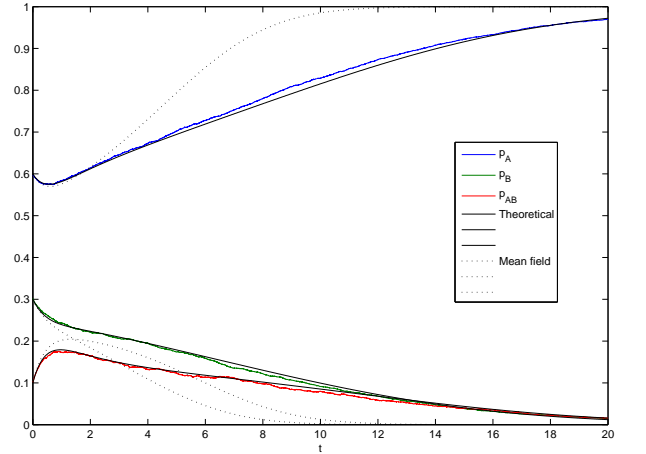}\\
  \caption{Fractions of A, B and AB nodes as function of time. The three color lines are the averages of 50 runs of NG (without committed agents) on ER network with $N=500$ and $<k>=5$.
  The black solid lines are solved from the ODE above with the same $<k>$. The black dotted lines are from the ODE using mean field assumption.}
  \label{figure:1}
\end{figure}

In this section, we show the numerical results of solving our ODEs by
Runge-Kutta method and compare the phase trajectories with those of
the basic mean field theory and also with the stochastic dynamical
trajectories of the simulated NG on random networks of varying
average degree. Fig.\ref{figure:1} shows the comparison between our
theoretical prediction (color lines) and the simulation on ER
networks (black solid lines). The dotted lines are theoretical
prediction by basic mean field approximation. We calculate the
evolution of the fractions of nodes with A, B and AB opinions
respectively and show that the prediction of the older basic mean
field approximation deviates from the simulations significantly
while that of the homogeneous pair approximation matches simulations
very well.

\begin{figure}[!hbp]
  \includegraphics[width=0.55\textwidth]{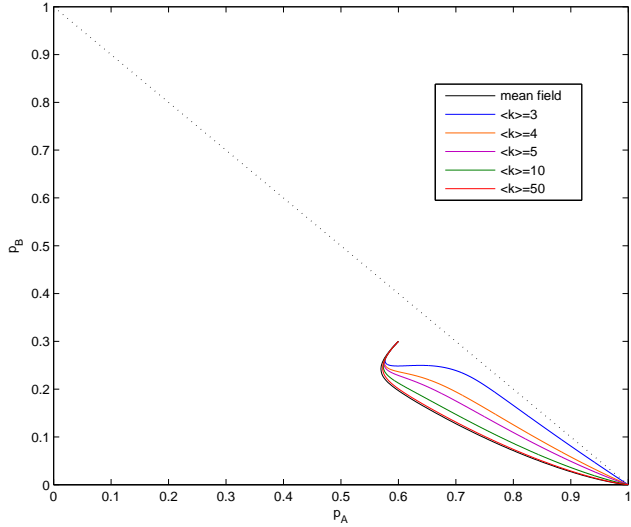}\\
  \caption{The trajectories of NG (no committed agents)solved from the ODE with different
  $<k>$ mapped onto 2D macrostate space. When $<k>\go \infty$, the trajectory tends to that of the mean field equation.
  When $<k>\go 1$, the trajectory get close to the line $p_{AB}=1-p_A-p_B=0$.}
  \label{figure:2}
\end{figure}

Fig.\ref{figure:2} shows the trajectories of the macrostate mapped onto two dimensional space ($p_A$,$p_B$), the black line is the trajectory predicted by the mean field
approximation. We find that when $<k>$ is large enough, say $50$, the homogeneous pair approximation is very close to the mean field approximation. When
$<k>$ decreases, the trajectory tends to the line $p_{AB}=1-p_A-p_B=0$, which means there are fewer nodes with mixed opinions than predicted by
the mean field. In this situation, opinions of neighbors are highly correlated forming the ``opinion blocks'', and mixed opinion (AB) nodes can only appear on the boundary between the ``A opinion block" and ``B opinion block".

\begin{figure}[!hbp]
  \includegraphics[width=0.55\textwidth]{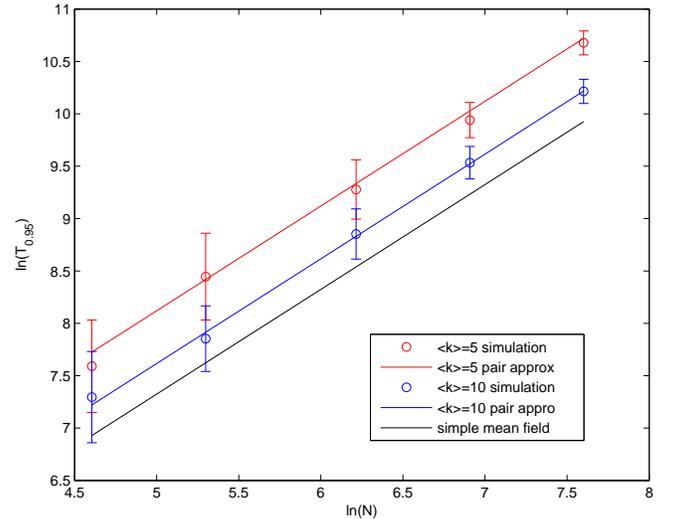}\\
  \caption{Comparison of $\eta$-consensus times $T_{0.95}$ of NG (no committed agent) between the simulation and theoretical prediction for different system sizes $N$ and average degrees $<k>$. The straight lines are from theoretical analysis under simple mean field assumption or pair approximation. The cycles and error bars show the means and the relative standard deviations of $T_{0.95}$ by simulation of dynamics}
  \label{figure:3}
\end{figure}

In the ODE models, it is hard to identify a proper cutoff for ``total consensus''. Therefore, to make a comparison between the theoretical prediction and the simulation, we consider $\eta$-consensus ($T_\eta$) which is the first time $p_A$ or $p_B$ achieves $\eta$. Fig.\ref{figure:3} shows the comparison of $T_{\eta}$ ($\eta=0.95$) for different system size $N$ and average degrees $<k>$. According to this figure, we find that when $N$ grows, the relative standard deviation of $T_{0.95}$ ($\Delta T_{0.95}/T_{0.95}\approx \Delta \ln(T_{0.95})$) decreases, which validates the pair approximation in the sense of thermodynamic limit. Further more, when $<k>$ grows, the pair approximation tends to the simple mean field assumption.

\section{Committed Agents}
In this section, we consider the asymmetric case of the NG on large random networks
with $p$ (fraction) committed agents (nodes that never change their
opinions) of opinion A. Initially, all the other nodes are of
opinion B. The main question considered here is
under what conditions it is possible for the committed nodes to persuade the
others and achieve a global consensus. Previous studies found there
is a robust critical value of $p$ called the tipping point. Above
this value, it is possible and the persuasion takes a short time,
while below this value, it is nearly impossible as it takes
exponentially long time with respect to the system sizes\cite{Xie11,Zhang11}.

Similar to what we did in the previous section, we derive the new
mean field ODE for the macrostate in the NG with committed agents,
although the macrostate now contains three more dimensions.
$\vec{L}=[L_{A-C},L_{B-C},L_{AB-C},L_{A-A},L_{A-B},L_{A-AB},L_{B-B},L_{B-AB},L_{AB-AB}]^T$,
where $C$ denotes the committed A opinion and $A$ itself denotes the
non-committed one. Hence we have a nine dimensional ODE which has
the same form as equation (\ref{eq2}), but with different details in
$D$ and $R$ given below:

{\small$$D=\left(
      \begin{array}{ccccccccc}
        0 & 0 & {3\over 4} & 0 & 0 & 0 & 0 & 0 & 0 \\
        0 & -{1\over 2} & 0 & 0 & 0 & 0 & 0 & 0 & 0 \\
        0 & {1\over 2} & -{3\over 4} & 0 & 0 & 0 & 0 & 0 & 0 \\
       0 & 0 & 0 &0 & 0 & {3\over 4} & 0 & 0 & {1\over 2} \\
       0 & 0 & 0 & 0 & -1 & 0 & 0 & 0 & 0 \\
       0 & 0 & 0 & 0 & {1\over 2} & -1 & 0 & 0 & 0 \\
       0 & 0 & 0 & 0 & 0 & 0 & 0 & {3\over 4} & {1\over 2} \\
       0 & 0 & 0 & 0 & {1\over 2} & 0 & 0 & -1 & 0 \\
       0 & 0 & 0 & 0 & 0 & {1\over 4} & 0 & {1\over 4} & -1 \\
      \end{array}
    \right),$$

   $$ Q_A=\left(
        \begin{array}{cccc}
         -1 & 0 & 0 & 0 \\
         0 & 0 & 0 & 0 \\
         1 & 0 & 0 & 0 \\
         0 & -1 & 0 & 0 \\
         0 & 0 & -1 & 0 \\
         0 & 1 & 0 & -1 \\
         0 & 0 & 0 & 0 \\
         0 & 0 & 1 & 0 \\
         0 & 0 & 0 & 1 \\
        \end{array}
      \right),
       \  \ Q_B=       \left( \begin{array}{cccc}
         0 & 0 & 0 & 0 \\
         -1 & 0 & 0 & 0 \\
         1 & 0 & 0 & 0 \\
         0 & 0 & 0 & 0 \\
         0 & -1 & 0 & 0 \\
         0 & 1 & 0 & 0 \\
         0 & 0 & -1 & 0 \\
         0 & 0 & 1 & -1 \\
         0 & 0 & 0 & 1 \\
        \end{array}\right),$$

\begin{eqnarray*}
R&=&( \vec{0},{1\over 2}Q_B \overrightarrow{P(\cdot|B)}, -{3\over 4}Q_A\overrightarrow{P(\cdot|AB)}, \vec{0}, {1\over 2}[Q_A \overrightarrow{P(\cdot|A)}+Q_B \overrightarrow{P(\cdot|B)}],\\
&&   Q_A [{1\over 4}\overrightarrow{P(\cdot|A)}-{3\over 4}\overrightarrow{P(\cdot|AB)}],\vec{0},Q_B[{1\over 4}\overrightarrow{P(\cdot|B)}-{3\over 4}\overrightarrow{P(\cdot|AB)}] ,\\
&& -(Q_A+Q_B)\overrightarrow{P(\cdot|AB)} ).\\
\end{eqnarray*}
}

\begin{figure}[!htbp]
  \includegraphics[width=0.55\textwidth]{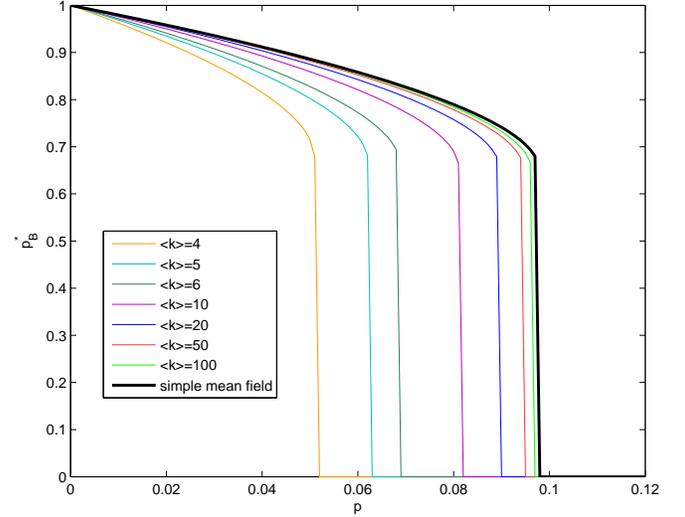}\\
  \caption{Fraction of B nodes of the stable point ($p_B^*$) as a function of the fraction of nodes committed to A (p).
  The color lines consist of stable points obtained by tracking the ODE of NG on ER for a long enough time. The black lines
  are the stable points solved from the mean field ODE.}
  \label{figure:4}
\end{figure}

\begin{figure}[!htbp]
  \includegraphics[width=0.55\textwidth]{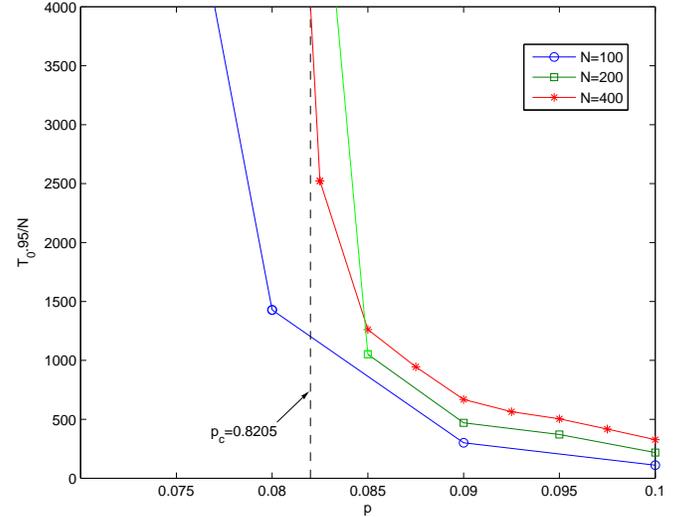}\\
  \caption{Normalized consensus time, $T_{0.95}/N$ around the tipping point $p_c=0.8205$ (vertical dash line) when $<k>=10$. Each data point is obtained by average of 100 runs of NG simulation with committed agents on ER network. The simulation stops when $T_{0.95}/N$ exceed $10^4$, since it almost never achieve consensus when $p<p_c$.}
  \label{figure:5}
\end{figure}

Finally, we show the change of the critical tipping fraction with
respect to the average degree $<k>$ of the underlying random
networks in Fig.\ref{figure:4}. Starting from the state that
$p_B=1-p$, the new ODE system will go to a stable state for which
$p_B=p_B^*$. $p_B^*$ is $0$ if the committed agents finally achieve
the global consensus. The sharp drop of each curve indicates the
tipping point transition with the corresponding $<k>$. Fig.\ref{figure:5} shows the normalized consensus time, $T_{0.95}/N$ around the tipping point $p_c$ for different system sizes. When $p>p_c$, $T_{0.95}/N$ is logarithmic with $N$; when $p<p_c$, $T_{0.95}/N$ grows very fast (since it takes to much time, we stop the simulation when $T_{0.95}/N$ exceeds $10^4$). Fig.\ref{figure:5} confirms the tipping point found in Fig.\ref{figure:4} is consistent with the transition point between the region of the logarithmic consensus time and exponential consensus time, and when the system size grows, the transition becomes sharper.

According to Fig.\ref{figure:4}, the tipping point shifts left when the average degree
$<k>$ decreases. This theoretical result confirms and replicate in
full without costly numerical simulations, the observed lowering of
the tipping fraction as a function of decreasing the average degree
of the underlying large random networks.

\newpage

\begin{acknowledgements}
This work was supported in part by the Army Research Laboratory
under Cooperative Agreement Number W911NF-09-2-0053, by the Army
Research Office Grants No. W911NF-09-1-0254 and W911NF-12-1-0546, and by the Office of
Naval Research Grant No. N00014-09-1-0607. The views and conclusions
contained in this document are those of the authors and should not
be interpreted as representing the official policies, either
expressed or implied, of the Army Research Laboratory or the U.S.
Government.

\end{acknowledgements}

\end{document}